\documentclass[twocolumn]{aastex631}

\begin{document}

\title{First constraints on Helium $^{+}{\rm He}^3$ evolution in $z=3-4$ using the 8.67GHz hyperfine transition}





\author[0000-0001-6324-1766]{Cathryn M. Trott}
\affiliation{International Centre for Radio Astronomy Research, Curtin University, Bentley, Australia}
\affiliation{ARC Centre of Excellence for All Sky Astrophysics in 3 Dimensions (ASTRO
3D), Bentley, Australia}

\author[0000-0002-6995-4131]{Randall B. Wayth}
\affiliation{International Centre for Radio Astronomy Research, Curtin University, Bentley, Australia}
\affiliation{ARC Centre of Excellence for All Sky Astrophysics in 3 Dimensions (ASTRO
3D), Bentley, Australia}

\begin{abstract}

We present the first constraints on the cross-correlation power spectrum of HeII ($^{+}{\rm He}^3$) signal strength using the redshifted 8.67GHz hyperfine transition between $z=2.9$ and $z=4.1$ and with interferometric data obtained from the public archive of the Australia Telescope Compact Array. 210 hours of observations of the primary calibrator source B1934-638 were extracted from data obtained with the telescope from 2014--2021, and coherently combined in a power spectrum pipeline to measure the HeII power across a range of spatial scales, and at three redshifts that span the period of Helium reionization. Our best limit places the brightness temperature fluctuation to be less than 557$\mu$K on spatial scales of 30 arcmin at $z=2.91$, and less than 755$\mu$K on scales of 30 arcmin at $z=4.14$ (2-sigma noise-limited). We measure a temperature of 489$\mu$K at $z=2.91$. ATCA's few antennas and persistent remaining RFI in the data prevent deeper integrations improving the results. This work is a proof of principle to demonstrate how this type of experiment can be undertaken to reach the 0.01--1$\mu$K level expected for the Helium signal at $z \sim 4$.
\end{abstract}


\section{Introduction} \label{sec:intro}
The reionization of neutral hydrogen in the first billion years of the Universe (HI$\rightarrow$HII) is a major phase transition in the intergalactic medium (IGM), with 75 percent of the IGM composed of hydrogen gas. This key period is being studied extensively through a number of observational tracers because it traces the formation of the first stars and galaxies \citep[likely responsible for the reionization;][]{furlanetto06}. One major observational tracer is the hyperfine transition of neutral hydrogen due to the energy-level splitting from the coupling of the spin states of the proton and electron, with a rest frequency of 1420~MHz (21~cm). The brightness temperature of this line encodes key information about the thermal and radiative state of the IGM, thereby indirectly probing the nature of the first stars and galaxies \citep{furlanetto06,koopmans15}. The ionisation energy of ground state hydrogen is 13.6eV, an energy available to be ionised by photons emitted in the ultraviolet part of the spectrum. The redshifted 21~cm transition is being pursued by many international experiments \citep{barry19,garsden21,hera,mertens20,gehlot19,trott20}.

Hydrogen combined at $z \approx 1100$ when the Universe had expanded and cooled sufficiently to bind the electron to the proton. At higher redshifts, $z \approx 6000$, Helium recombined (HeIII $\rightarrow$ HeII), when its inner electron, with a binding energy of 54.4eV, recombined \citep{switzer08}. Helium-4 comprises almost 24 percent, by number, of the IGM after recombination, but Helium-3, which has a non-zero magnetic dipole moment that can produce hyperfine splitting, has a substantially smaller abundance \citep[1 part in 10$^5$,][]{kneller04}. The first electron of neutral helium has a similar binding energy as hydrogen and is expected to reionize at similar redshifts to hydrogen ($z > 6$), however the reionization of singly- to double-ionised helium is expected to occur much later, when AGN can provide sufficient high energy photons to unbind the 54.4eV second electron. This HeII to HeIII process therefore is the second major reionization period of the Universe, liberating further electrons into the IGM. Like hydrogen, the hydrogenic single-electron $^{+}{\rm He}^3$ ion has a hyperfine splitting of energy states, emitting a 8.67~GHz photon (3.5~cm, rest).

There is indirect evidence for Helium reionization being complete by $z \sim3$, including optical depth to HeII \citep{worseck11,worseck16}, increased IGM temperature \citep{lidz10,makan21}, and hardness of background radiation from metal ionization potentials \citep{turner16,morrison19}. Fast Radio Bursts (FRBs) also have the potential to probe this era because the spectral dispersion of their signals is linearly proportional to the line-of-sight electron component of the IGM \citep{caleb19}.

\citet{worseck11,worseck16} have studied the spectra of $z > 3.5$ AGN for evidence of Gunn-Peterson absorption of helium Ly-$\alpha$ Forest of the redshifted 304 Angstrom line, reporting that the IGM was highly-ionized by $z=3.4$ due to large transmission regions, and high variance between sightlines. Like the hydrogen Ly-$\alpha$ Forest, this probe provides detailed information along skewers through the IGM, but is less sensitive to the large-scale signal evolution. Similarly, the large optical depth to the Ly-$\alpha$ line (and the saturation of the line at low neutral fraction) makes this probe sensitive primarily to the end of reionization. \citet{mcquinn10} considered the HeI Ly-$\alpha$ forest, with rest wavelength of 584 Angstrom, as a tracer of the reionization of HeII. Unlike the HeII forest, which saturates easily and has a shorter wavelength, this tracer is weakly absorbed, providing the potential for a quantitative study of singly-ionised helium. 

Radio observation of the reionization of helium is via the radio hyperfine line, line-of-sight absorption to background AGN (Ly-$\alpha$ Forest - 304 Angstrom line), and indirectly in the IGM electron density (e.g., through observations of the dispersion measure - redshift relation with Fast Radio Bursts and other high-redshift transients). Optical tracers can probe Helium-4 because they rely on electron transitions. To-date, there has been no detection of the Helium-3 hyperfine transition, or attempts to undertake this experiment. Observationally, its small optical depth relative to hydrogen (due primarily to its very small primordial abundance) is balanced somewhat by the lower system temperature and reduced radio foreground at higher frequencies. A number of studies have predicted theoretically \citep{mcquinn09,bagla09}, or simulated \citep{khullar20}, the helium hyperfine signal in $z=3-4$, predicting temperature fluctuations spanning 0.1--50$\mu$K on different scales. \citet{khullar20} considered the temperature fluctuations of HeII around QSOs in simulations at higher redshifts, commensurate with hydrogen reionization.

Detection of intergalactic $^{+}{\rm He}^3$ requires a contrast between the resonant signal temperature and the CMB temperature. \citet{mcquinn09} argues that both weak Wouthysen-Field coupling and small collisional coupling in the IGM would make the temperature contrast small and the signal very difficult to measure. Instead, they consider three avenues for using $^{+}{\rm He}^3$ to understand the evolution of the Universe. Firstly, because absorption of continuum light to a background source by HeII is proportional to the source flux density (and not the CMB), identifying and performing spectral absorption measurements on $z>3.5$ QSOs would be possible. Secondly, at higher redshift, when HeI is transitioning to HeII, first producing the 8.67GHz hyperfine signal, absorption would probe HeI-HeII reionization. Finally, HeII that is self-shielded in halos with sufficient density for collisional coupling to produce an emission signal, could be used to trace the biased signal evolution (similar to post-reionization hydrogen).

In this paper, we aim to place constraints on the spherically-averaged (three-dimensional) power spectrum of brightness temperature fluctuations of singly-ionised Helium-3 over $z=2.9-4.2$. This work acts as a demonstration of how to undertake this experiment with real telescope data. In Section \ref{sec:methods} we introduce the methods, including theoretical predictions and observations; in Section \ref{sec:analysis} the power spectrum methodology is introduced; in Section \ref{sec:results} the results are presented, before being discussed in Section \ref{sec:discussion}.

\section{Methods} \label{sec:methods}
\subsection{Theoretical predictions and observational approaches}
The optical depth to $^3$He$^+$ is given by \citet{mcquinn09}:
\begin{equation}
    \tau_{\rm He^+} = \frac{c^3\hbar A_{10}}{16k_BT_s\nu^2_{10}} \frac{n_{\rm {^3}He^+}}{(1+z)dv/dy},
\end{equation}
where $A_{10}$ is the spontaneous emission rate, $\nu_{10}$=8666MHz is the rest frequency, $n_{\rm {^3}He^+}$ is the Helium-3 gas density, which is computed as a fractional abundance of the hydrogen density (1.04 $\times$ 10$^{-5}$), and $dv/dy$ is the proper velocity per unit conformal distance. Although the helium fraction to hydrogen is small, the spontaneous emission rate is 670 times larger, offsetting some of the reduction. The spin temperature, $T_s$, is set by the relative occupancy of the two energy states, and like hydrogen, is set by the thermal, collisional and radiative properties of the medium. The differential brightness temperature of the emission (temperature relative to the CMB), is then given by \citep{bagla09,furlanetto06}:
\begin{eqnarray}
    \delta{T}_b \simeq &18\mu\textrm{K} \,\, x_{\rm HeII} \left( 1-\frac{T_{CMB}}{T_s} \right) \left( \frac{1+\delta}{10} \right) \left( \frac{1+z}{9} \right)^2  \nonumber \\
    & \times \left( \frac{n_{3,He}/n_H}{1.1 \times 10^{-5}} \right) \left[ \frac{H(z)}{(1+z)(dv_\parallel /dr_\parallel)} \right].
    \label{eqn:deltat}
\end{eqnarray}
During hydrogen reionization ($z=6-10$), brightness temperature fluctuations in Helium-3 are driven by spin temperature coupling to the gas temperature. \citet{bagla09} suggest that helium fluctuations at the time of hydrogen reionization are driven by density inhomogeneities and coupling to the gas kinetic temperature, which can be $10^4$K in ionized regions. In the hydrogen post-reionization era, they propose that high-energy photons from AGN can reionize Helium-3, and that brightness temperature fluctuations will be dominated by ionization status and gas kinetic temperature inhomogeneities. 

Equation \ref{eqn:deltat} shows that a non-zero emission signal relies on $T_{\rm CMB} \neq T_s$. At $z=3.6$, \citet{mcquinn09} argue that the low collisional coupling in the IGM and weak radiative spin coupling prevents $T_s$ from coupling to the gas kinetic temperature and instead keeps it coupled to the CMB temperature. Observation of emission signal is therefore only feasible for HeII that is self-shielded from reionization in halos, where the density is such for collisional coupling to increase the spin temperature. These overdense regions ($\Delta_b\geq 100, n_e \geq 3 \times 10^{-6} cm^{-3}$) may comprise $X=1-10\%$ of the gas mass density. In IllustrisTNG simulations, \citet{martizzi19} found that 1--15\% of gas resided at these densities from $z=4-2$. In this scenario, the helium signal traces the halo distribution with an order-unity bias, $b$, giving a nominal power of:
\begin{eqnarray}
    P(k) &\simeq& \left(18\mu\textrm{K} \left( \frac{1+\delta}{10} \right) \left( \frac{1+z}{9} \right)^2\right)^2 X^2 b^2 P_m(k) \nonumber\\
    &=& (0.6\mu\textrm{K})^2 X^2 b^2 P_m(k) \,\, h^{-3}\textrm{Mpc}^3
    \label{eqn:expected}
\end{eqnarray}
where $P_m(k)$ is the matter power spectrum, and we have assumed $z=4$ and $\delta=0$. We use the CAMB software\footnote{https://camb.info/} \citep{lewis02,lewis13}, to generate the matter power spectrum between $z=2.9-4.1$.

At $z \simeq 3-4$, the redshifted hyperfine line is observed at 1700--2200 MHz, in contrast to hydrogen's 21cm line being redshifted to 100--200 MHz. Observationally, the sky temperature at higher frequencies (which dominates the system temperature at low frequencies), is substantially lower, yielding lower radiometric noise ($T \propto \nu^{-2.6}$). Other key factors in determining the utility of observations are the field-of-view ($\propto \lambda$), the instantaneous bandwidth (for continuum foreground fitting and removal) and spectral resolution (for matching ionization regions), and the amplitude of ionospheric refraction ($\propto \lambda^2$).

Simulations of \citet{compostella13} and \citet{mcquinn09_2} show that IGM kinetic temperature and ionization fraction fluctuations can be observed on spatial scales of 1--50 cMpc at $z=3.5$, corresponding to 1--20 arcmin angular scales and 1--30 MHz spectral scales from an observational standpoint. Any experiment to attempt the intergalactic signal would be suited to an array with baselines of $x = u\lambda =  \lambda/\Delta\theta \simeq 30-500$ metres (at $\lambda=17$ cm), and a dish aperture of $d = \lambda/$FOV $< 30$ metres. These parameters are well-matched to the ATCA, and provide the starting point for our analysis.

\subsection{Observations}
The Australia Telescope Compact Array\footnote{https://www.narrabri.atnf.csiro.au/} (ATCA) is a 6-element dish-based interferometer that is mostly operated in an East-West configuration, at the Paul Wild Observatory near Narrabri NSW. Each dish has a diameter of 22~m, and is fed by multiple interchangeable feeds, operating at centimeter and millimetre wavelengths. The CABB backend, installed in 2009, allows for an instantaneous 2048 MHz of bandwidth \citep{wilson11}. For our experiment, we limit data to those observed in the 1.1-3.1~GHz CABB band, with 1~MHz spectral resolution over 2048 channels. The ATCA operates in a number of array configurations. For this work, the archive was searched for all 6km, 1.5km and 750m configurations, with array minimum and maximum baselines spanning 30.6~m to 6000~m. The shorter of these baselines, and the spectral resolution and bandwidth of the CABB, are well-suited for an interferometric helium experiment. What is less well-suited, is the low sensitivity afforded by the small number of antennas. With the 6th antenna (fixed at a minimum baseline of 3~km from the others) effectively unusable for this science, there are only 10 baselines available for integration. This will limit the sensitivity of the experiment compared with another array with better $uv$-coverage and sensitivity.

Due to the low sensitivity of the ATCA, and the desire to undertake a proof-of-principle experiment, we chose to use public data from the extensive ATCA archive, rather than propose for new multi-year observations. To undertake a coherent experiment (where the noise can be reduced most effectively by observing the same patch of sky for the full observation), a field is required where the ATCA has spent substantial observing time, and preferably with a simple foreground source that can be easily removed. The primary calibrator source, B1934-638, fits both of these criteria: it has been one of the most-observed calibrator sources for data over the CABB lifetime, and it is a well-modelled unresolved source, with a single power law spectrum and a flux density of $\sim$12~Jy at 1500~MHz.

The ATCA archive\footnote{https://atoa.atnf.csiro.au/} was searched for observations containing source B1934-638, using the CABB backend over 1.1--3.1~GHz with 1~MHz spectral resolution, for all 6km, 1.5km and 750m baseline configurations, and for dates spanning 2014--2022 (public data). These individual observations were inspected manually to extract those with $>$30 minutes continuous observation of the source. This provided a set of longer observations to reduce the number of observations to be flagged. The search yielded 410 hours of data.

\subsection{Data analysis}\label{sec:analysis}
We used the data reduction software \textsc{Miriad} \citep{miriad} to flag and calibrate data for further analysis. The CABB backend outputs data across two IFs. For the data used in this project, the channels were listed in reverse order (decreasing frequency). After visual inspection of a few datasets using uvspec, it was clear that there were bands of persistent RFI that affect all CABB data. These were flagged across all datasets using uvflag. In addition, the top and bottom of the band were flagged due to large regions of RFI, and the desire to analyse data outside of the local 21cm band (1420~MHz) and below 2.2~GHz. The first 400, and final 827 channels were flagged, leaving 650 channels spanning 1676--2325~MHz ($z=4.2-2.7$). Each dataset was then manually inspected using \textsc{uvspec}, with auto- and cross-correlations visualised after averaging each 60 minutes of data. RFI identified by visual inspection were manually flagged using \textsc{uvflag}. In addition poorly-performing baselines or antennas were also flagged. After the first round of flagging, data were calibrated using \textsc{mfcal}. Because B1934-638 is the primary calibrator for ATCA, no other field data were required. After calibration, datasets were inspected again with \textsc{uvspec}. All data that appeared RFI-free and showed cross-correlations consistent with expectations (flat $\sim$12~Jy continuum source) were retained for further analysis. After rejection of poorly-calibrated observations, there were 210 hours of data.

The foreground source, B1934-638, is well-fitted by a first-order polynomial. Fitting across the full 650~MHz does not affect the Helium-3 structures of interest to this work, due to the very broad backend. First-order fits were subtracted from the visibilities using the uvlin procedure. Residual data and weights were written to UVFITS format using the \textsc{Miriad} \textsc{FITS} procedure for the XX and YY polarisations, and all baselines, with a temporal and spectral resolution of 10~seconds and 1~MHz, respectively.

\subsection{Power spectrum estimation}
For hydrogen reionization, there are many power spectrum estimation pipelines available. For the Murchison Widefield Array EoR project, we employ the CHIPS software to estimate the spherically-averaged power spectrum \citep{trottchips}. CHIPS natively reads the uvfits files produced by \textsc{Miriad}. We adjusted the software to create CHeIIPS, a package that uses the same methodology but with parameters updated to match that for the Helium-3 reionization experiment (e.g., gridding kernel size, uv-plane resolution).

Full details of the methodology are available in \citet{trottchips}, and is briefly reproduced here. Channelised calibrated visibilities are cumulatively gridded onto a three-dimensional uv-plane grid ($u,v,\nu$) as complex doubles, along with a separate weights grid that accumulates the gridding kernel. The kernel would optimally be set to the Fourier Transform of the telescope primary beam; for CHIPS and CHeIIPS, a matched-size 2D Blackman-Harris window is employed due to its better sidelobe suppression performance. In \citet{trottchips}, this choice was shown to not produce signal loss. Data are split into two datasets, interleaved in time, where each consecutive 10~second timestep is assigned alternately to even and odd grids. This allows for the data cross power spectrum to be computed, whereby noise power that would be present due to squaring of a quantity is absent (noise being uncorrelated between timesteps). This choice comes at the expense of slightly higher noise uncertainty.

In the first stage of analysis, data were coherently gridded into 20 sets of 10 hours each. This allowed for inspection of individual sets of power spectra to identify any poorly-behaved observations that had passed calibration assessment. After coherent gridding, the data are normalised by the weights to produce the averaged gridded data:
\begin{equation}
    \tilde{V}(u,v,\nu) = \frac{\displaystyle\sum_i V_iW_i}{\displaystyle\sum_i W_i},
\end{equation}
where $W_i$ is the visibility weight multiplied by the gridding kernel.

The final frequency transform to Fourier Space ($\nu-\eta$) is performed for each $uv$-cell. For CHIPS, where the residual foregrounds are bright and complex, a Blackman-Harris window function is employed to reduce foreground sidelobe contamination. This is undertaken at the cost of a broader main lobe in $\eta$ space. For these Helium-3 data, the foreground source has been effectively removed by fitting the first-order polynomial, and no spectral window function is required. 

After spectral transform via a DFT, the final $(u,v,\eta)$ data and weights cubes are used for power spectrum analysis. Transform across the full 650~MHz is not appropriate, because the signal evolves and is not ergodic. In addition, the many missing spectral channels due to flagging of persistent RFI leave gaps in the data that produce very poor power spectral results. As such, we choose bands of gridded data that contain no missing data, and limit the line-of-sight transform to a maximum of 100~MHz ($\Delta{z} \simeq 0.25$). Eight over-lapping subbands were identified that met these criteria. These are described in Table \ref{table:bands}.
\begin{table}
\centering
\begin{tabular}{|c||c|c|}
\hline 
$\nu_{\rm low}$ (MHz) & $z_{\rm cent}$ & $N_{\rm ch}$ \\ 
\hline \hline 
1676 & 4.17 & 60\\
\textit{1686} & 4.14 & 80\\
1781 & 3.86 & 54\\
1902 & 3.55 & 96\\
1933 & 3.48 & 96\\
\textit{1974} & 3.39 & 54\\
2110 & 3.10 & 42\\
\textit{2216} & 2.91 & 60\\
\hline
\end{tabular}
\vspace{0.1cm}
\caption{Starting frequencies, central redshift and number of spectral channels for each subband selected from the data. Italicised subbands were used for the final analysis.}\label{table:bands}
\end{table} 

Cylindrically- and spherically-averaged power spectra are produced by squaring the gridded, transformed visibilities and incoherently averaging them using the gridded weight data. The cylindrical transform plots angular ($k_\bot = \sqrt{k_x^2+k_y^2}$, $k_x \propto u$) and line-of-sight ($k_\parallel \propto \eta$) modes separately:
\begin{equation}
    P(k_\bot,k_\parallel) = \frac{\displaystyle\sum_{i \in k} V_i^\ast V_i W_i^2}{\displaystyle\sum_{i \in k} W_i^2} \,\,\text{Jy}^2 \text{Hz}^2,
\end{equation}
where $W_i$ is the summed weights over all frequency channels. The power is transformed from observational units of Jy$^2$ Hz$^2$ to cosmological units of mK$^2$ h$^{-3}$ Mpc$^3$ using standard transforms and the transverse comoving distance at the sub-band centres. Finally, the dimensionless power spectrum is computed as:
\begin{equation}
    \Delta^2(k) = \frac{k^3 P(k)}{2\pi^2} \,\, \mu{\rm K}^2,
\end{equation}
where a final conversion from mK to $\mu$K is performed.
In one dimension, we display the square-root of this quantity, equivalent to the spectrum of temperature fluctuations.

\section{Results} \label{sec:results}
The 2D and 1D power spectra were inspected for each subset of 10 hours of observations, for each sub-band described in Table \ref{table:bands}. Most subsets displayed reasonable results, where the power were consistent with thermal noise across most wavemodes. One subset showed clear systematic contamination and was removed from the final coherent average. The remaining subsets were coherently averaged to $\simeq$190 hours of data, except for the $z=4.14$ subband, where only 120 hours of data showed good behaviour.

\begin{figure}[ht!]
\plotone{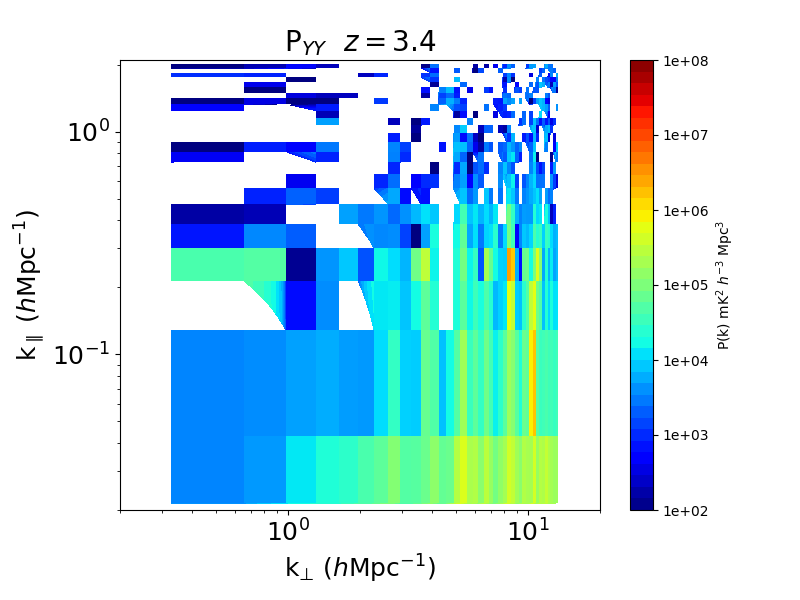}
\caption{Cylindrically-averaged power spectrum at $z=3.4$ over 54~MHz of bandwidth, and the YY polarisation. Missing data denote negative cross-power, highlighting noise-like regions. This figure has units of mK$^2$ $h^{-3}$ Mpc$^3$.
\label{fig:2D_3pt4_YY}}
\end{figure}

The eight individual subbands showed different systematic behaviour. Three that spanned the full band displayed the cleanest data, with most modes consistent with thermal noise. These subbands are italicised in Table \label{table:bands}, and correspond to power at $z=2.91, 3.39, 4.14$. These were processed to the final power spectra. Figure \ref{fig:2D_3pt4_YY} displays a 2D power spectrum at $z=3.39$ in units of mK$^2$ h$^{-3}$ Mpc$^3$. The DC ($k_\parallel=0$) mode has been omitted. The data show positive and negative modes, consistent with thermal noise when cross power spectra are considered. The positive modes at high $k_\bot$, low $k_\parallel$ show residual foreground power that has not been removed by the linear fitting to the visibilities.

\begin{figure*}[ht!]
\includegraphics[width=0.55\textwidth]{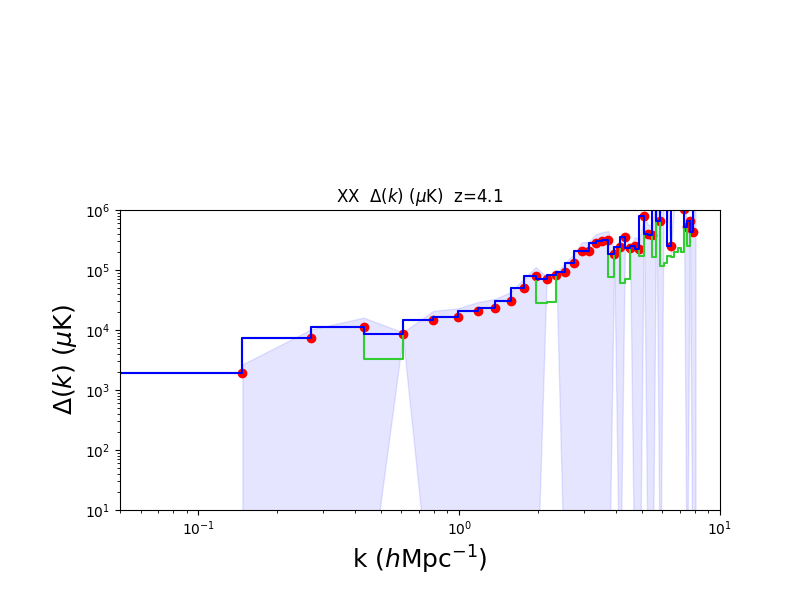}
\hspace{-1cm}
\includegraphics[width=0.55\textwidth]{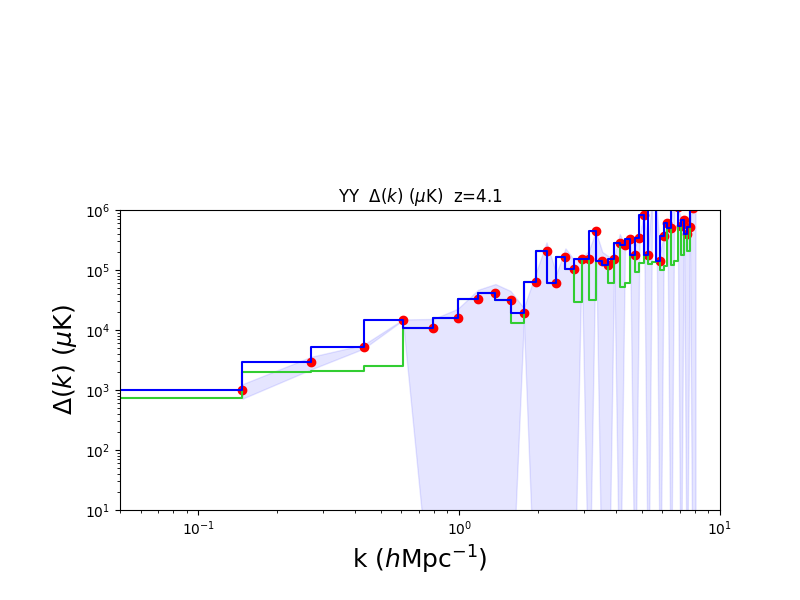}\\
\includegraphics[width=0.55\textwidth]{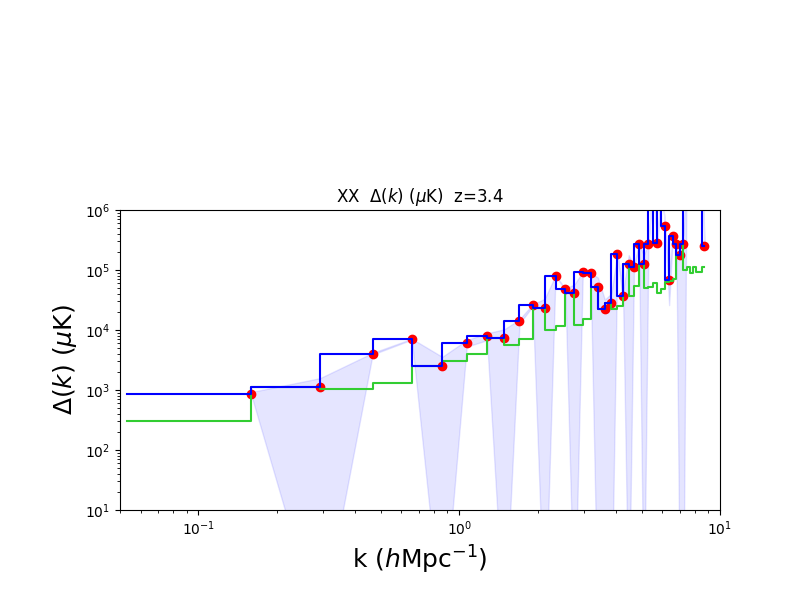}
\hspace{-1cm}
\includegraphics[width=0.55\textwidth]{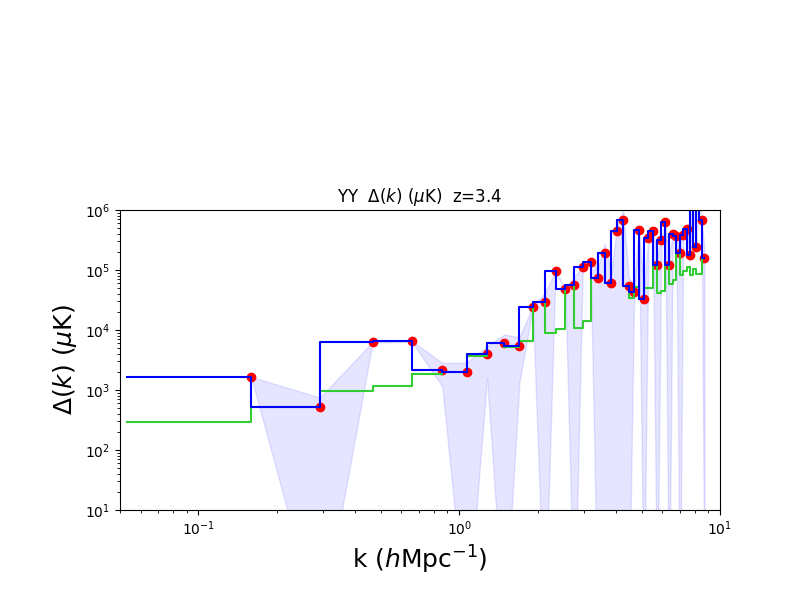}\\
\includegraphics[width=0.55\textwidth]{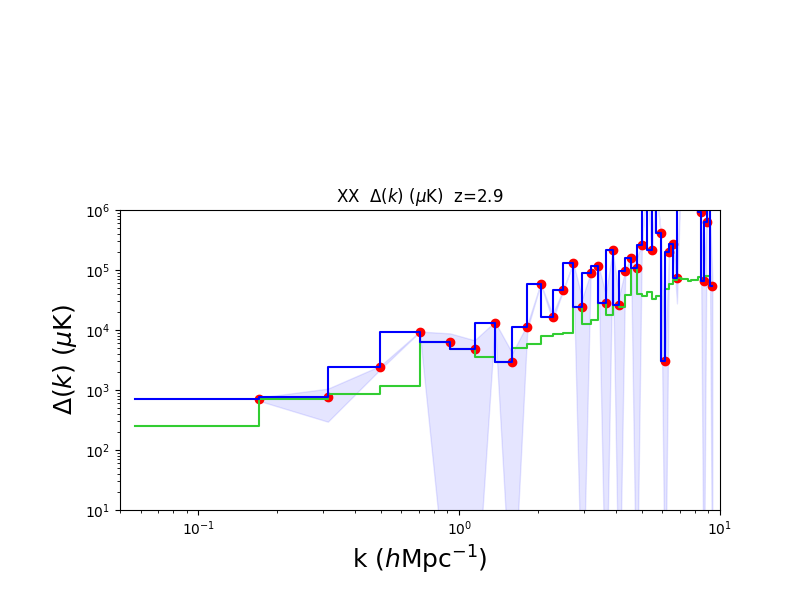}
\hspace{-1cm}
\includegraphics[width=0.55\textwidth]{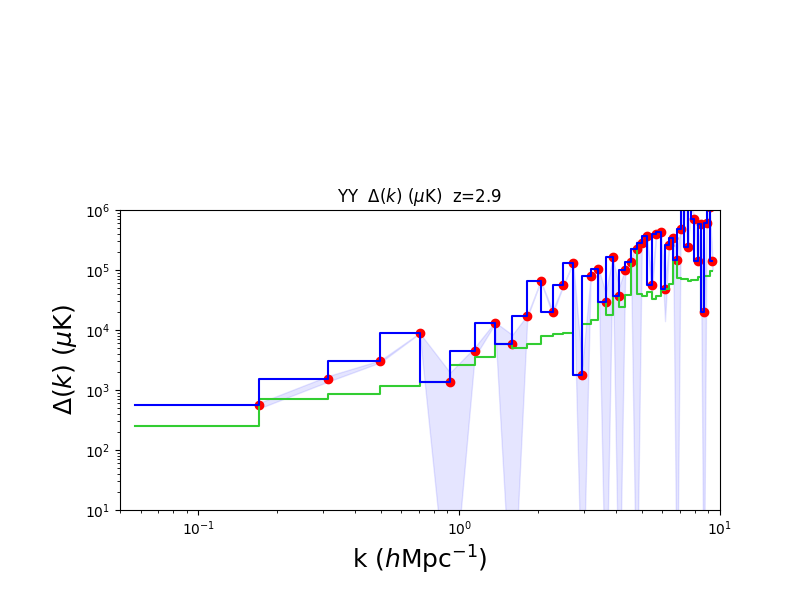}
\caption{Spherically-averaged temperature fluctuation ($\mu$K) as a function of wavenumber for $z=4.1, 3.4, 2.9$ (top-to-bottom) across 80~MHz of bandwidth and XX (left) and YY (right) polarisations. The blue shaded region denotes the 2$\sigma$ thermal noise with noise-dominated regions showing shading to the bottom of the plot. The red points and blue line denote the measured power, while the green line denotes the estimated 2$\sigma$ noise level.
\label{fig:1D_4pt1}}
\end{figure*}

\begin{figure}
    \centering
    \includegraphics[width=0.48\textwidth]{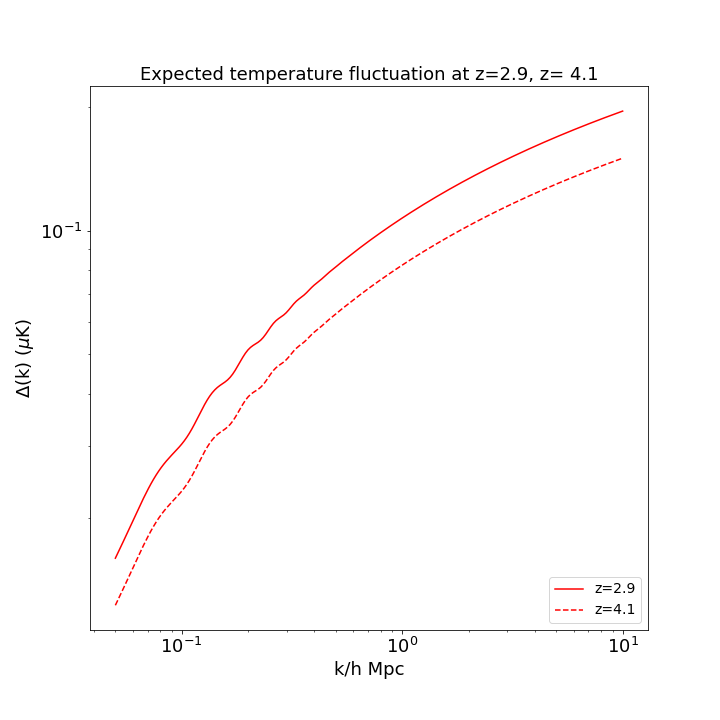}
    \caption{The expected temperature fluctuation of the shielded model from Equation \ref{eqn:expected}, for $z=2.9, 4.1$ and assuming that 10\% of gas is collisionally-coupled.}
    \label{fig:expected}
\end{figure}

The three subbands of data are then spherically-averaged to 1D, for each polarization, and square-rooted to produce the spectrum of temperature fluctuations shown in Figure \ref{fig:1D_4pt1}.
In each, the blue shaded region denotes the 2$\sigma$ thermal noise with noise-dominated regions showing shading to the bottom of the plot. The red points and blue line denote the measured power, while the green line denotes the estimated 2$\sigma$ noise level. Many modes are consistent with thermal noise. The best measured temperature is at $z=2.91$ with $\Delta{(k)}$ = 489$\mu$K, and a 2$\sigma$ upper limit of 557$\mu$K. Table \ref{table:results} displays the measured and upper limit temperature fluctuations for the three subbands and their angular scale. For all redshifts, the YY power spectrum yielded cleaner results, and are reported here. The XX polarisation results are only marginally poorer, and this likely points to localised RFI that affects one polarisation more than the other.
\begin{table}
\centering
\begin{tabular}{|c||c|c|c|c|}
\hline 
$z_{\rm cent}$ & Meas. $\Delta(k)$ $\mu$K & 2$\sigma$ limit & k (h$^{-1}$ Mpc) & $\Delta\theta$ (') \\ 
\hline \hline 
4.14 & 534 & 757 & 0.15 & 29\\
3.39 & 689 & 760 & 0.16 & 29\\
2.91 & 489 & 557 & 0.17 & 29 \\
\hline
\end{tabular}
\vspace{0.1cm}
\caption{Measured temperature and 2$\sigma$ upper limits for each of the three subbands and the YY polarization, and the angular mode of the lowest temperature.}\label{table:results}
\end{table} 
In each case, the angular mode corresponds to the shortest baseline available to the ATCA (30.6~m), which is typically occupied by a single baseline, and therefore has poor sensitivity. For comparison, Figure \ref{fig:expected} shows the expected temperature fluctuation of the shielded model from Equation \ref{eqn:expected}, for $z=2.9, 4.1$ and assuming that 10\% of gas is collisionally-coupled. The broad level is $\sim0.01- 0.1\mu$K, which is 3--4 orders of magnitude below the measurements, demonstrating the difficulty of the experiment.

\section{Discussion}\label{sec:discussion}
This work is an attempt to place upper limits on the power spectrum of temperature fluctuations for Helium-3, and provides an approach to how to undertake such an experiment. Despite not having the sensitivity to theoretically detect the signal, our data exhibit two important characteristics that currently impede interferometric 21~cm experiments pursuing hydrogen reionization: (1) A single clean and simple foreground source that can be removed effectively, due to the wide instantaneous bandwidth; (2) sufficiently-low residual RFI that the data are mostly noise-limited at 200 hours.

The accuracy of the subtraction of B1934-638 is sufficient at the 200 hour level, but there are indications that both a first- and second-order polynomial fit leaves residuals that are not consistent with zero-mean noise in some parts of the band. The residual signal in a 1~MHz channel should not exceed 1~$\mu$Jy, according to \citet{mcquinn09}, requiring a dynamic range of 10$^7$ for a 12~Jy source. If the source can be fitted by a low-order polynomial, then the full band can be used for foreground estimation and subtraction, providing sufficient SNR. E.g., across 650 channels and fitting for a single parameter, 200 hours with all 15 ATCA baselines yields a dynamic range of 2$\times$10$^6$. Increasing the order of the polynomial fit increases the chances of signal loss, by introducing spectral structure that has a non-zero projection onto the spectral Helium-3 structure (i.e., complex spectral structure may be due to Helium-3, and not the foreground source), whereas a first-order polynomial over 650~MHz will not impact the underlying Helium-3 structure (any global Helium-3 signal evolution has already been removed by omitting the auto-correlations).

\begin{figure}[ht!]
\plotone{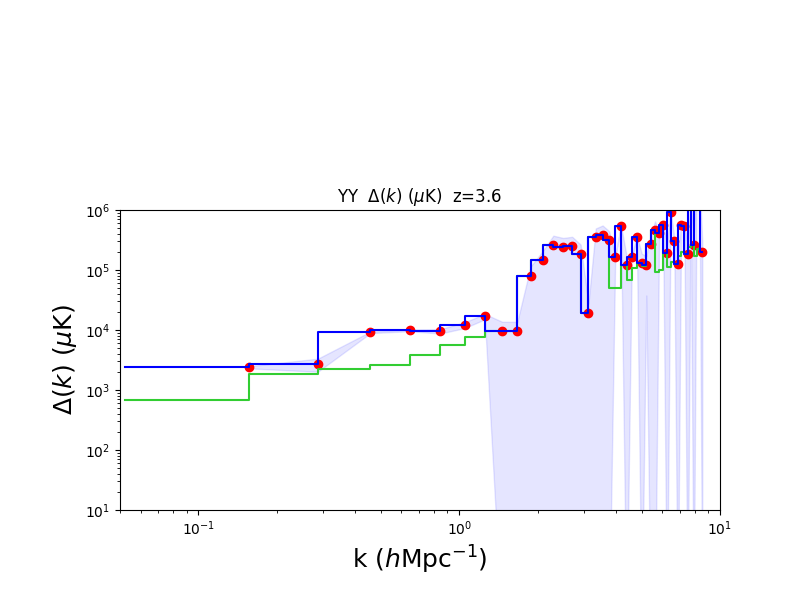}
\caption{Spherically-averaged temperature fluctuation ($\mu$K) as a function of wavenumber for $z=3.6$ across 96~MHz of bandwidth and the YY polarisation. These data are affected by residual spectral structure from RFI or the foreground fitting leading to poor performance at low $k$.
\label{fig:baddata}}
\end{figure}

Residual RFI and spectral structure does affect a lot of the band; hence why only three out of the eight bands were used for the final analysis. Figure \ref{fig:baddata} displays YY power spectra for a poorly-performing subband of 96 channels near $z=3.6$. For these data, there is residual spectral structure across the subband visible in the gridded visibility spectra, leading to increased leakage from foregrounds modes.

It may be noted that the auto-correlation data may be used to study the global temperature evolution of the Helium-3 signal. With a single power-law continuum source as the only bright foreground in the field, a broad non-power law evolution of the spectrum may indicate the global reionization of Helium \citep[similar to that undertaken with the EDGES experiment for hydrogen;][]{bowman18}. However, inspection of the auto-correlations from ATCA show high-amplitude oscillations, consistent with reflections off structures within the dishes. The systematics in these data were not deemed able to be modelled sufficiently to undertake this experiment. Similarly, one may perform a Helium-3 absorption study with the cross-correlation spectra, \textit{if} B1934-638 were a background, high-redshift source (instead, it is $z=0.12$). \citet{mcquinn09} predicted that absorption against a background continuum source provided the highest detectability, because the absorption flux density is dictated by the strength of the source and the optical depth. Given the lack of expected emission signal from the IGM, and the low signal expectation from the self-shielded gas in halos, the absorption study may be the best observational approach to tackle.

ATCA is well-suited to undertake this experiment, due to its wide CABB backend and small field-of-view, but is ultimately hampered by its few baselines and lack of sensitivity. Prospects for detection with other current and future telescopes is dependent on: (1) overall sensitivity; (2) field-of-view (too small and the results are sample variance-limited, and too large and the foregrounds become complicated); (3) low-RFI environment to retain spectral smoothness and afford bands without flagged channels; (4) short baselines to access large-scale structures; (5) a broad, instantaneous bandwidth to measure the full signal evolution and for accurate foreground removal. Few telescopes have all of these properties; in particular the wide-band backend system.

SKA-Mid is an obvious instrument for this experiment. Band 3, which is not proposed to be available initially\footnote{http://www.skao.int}, spans 1400~MHz instantaneously across 1650--3050~MHz. Bands 1 and 2 would probe Helium at higher redshift. Using the SKA-Mid baseline distribution \citep{braun}, there will be $\simeq$90 baselines with physical lengths shorter than 50~m. With this configuration, a 100-hour, 100~MHz bandwidth experiment should yield noise levels of 1$\mu$K at $k=0.14h$Mpc$^{-1}$, in the absence of RFI and other spectral systematics. The precursor MeerKAT array has $\simeq$10 baselines shorter than 50~m (minimum 29~m), and with an existing wideband S-band receiver, should reach 1$\simeq \mu$K noise level after 1000~hours. Other facilities with good $uv$-coverage have narrower fractional bandwidth at these frequencies, which will hinder their ability to perform accurate foreground subtraction without the potential for signal loss (e.g., VLA). Higher-redshift $^+$He$^3$, as explored by \citet{khullar20}, is accessible at lower frequencies, but there are few radio interferometers with feeds that access 800--1200~MHz. For gas around $z=5.1$ (1420~MHz), the signal will be confused with local neutral hydrogen, making its observation almost impossible.

\section{Conclusions}
We attempted a measurement of the spherically-averaged power spectrum of $^+$He$^3$ (single-ionised Helium-3) using the 3.5~cm (8.67~GHz) hyperfine line, at $z=2.9-4.1$ using 190 hours of publicly-available data from the ATCA archive. After RFI flagging, identification of spectral bands with no missing data, and a simple foreground subtraction, power spectra were found to be noise-limited across many angular wavemodes. The upper limits on the power are not cosmologically-interesting, being 3--4 orders of magnitude larger in temperature than theoretical expectations, but is the first attempt to measure this signal. This work demonstrates the potential for this experiment to yield improved results, which would be cosmologically-relevant, with a telescope with higher sensitivity.

\begin{acknowledgments}
The authors would like to specifically thank the referee for their detailed comments and correction of our understanding of the theory. This clarification has made a significant improvement to the manuscript. The authors would also like to thank Jishnu Thekkeppattu and Jamie Stevens for helpful discussions. This research was partly supported by the Australian Research Council Centre of Excellence for All Sky Astrophysics in 3 Dimensions (ASTRO 3D), through project number CE170100013. CMT was supported by an ARC Future Fellowship under grant FT180100321.
The International Centre for Radio Astronomy Research (ICRAR) is a Joint Venture of Curtin University and The University of Western Australia, funded by the Western Australian State government.
\end{acknowledgments}

%

\vspace{5mm}
\facilities{Australia Telescope Compact Array}


\software{\textsc{Miriad}, https://www.atnf.csiro.au/computing/software/miriad/ \citep{miriad}}


\end{document}